\documentclass[twocolumn,pra,showpacs]{revtex4}
\usepackage{amsfonts}
\usepackage{amsmath}

\begin{document}

\title{Non-Markovian master equations from entanglement with stationary
unobserved degrees of freedom}
\author{Adri\'{a}n A. Budini$\,^{1}$ and Henning Schomerus$\,^{1,2}$}
\affiliation{$\,^{1}$Max Planck Institute for the Physics of Complex Systems, N{\"o}%
thnitzer Stra{\ss }e 38, 01187 Dresden, Germany\\
$\,^{2}$Department of Physics, Lancaster University, Lancaster LA1 4YB, UK}
\date{\today}

\begin{abstract}
We deduce a class of non-Markovian completely positive master equations
which describe a system in a composite bipartite environment, consisting of
a Markovian reservoir and additional stationary unobserved degrees of
freedom that modulate the dissipative coupling. The entanglement-induced
memory effects can persist for arbitrary long times and affect the
relaxation to equilibrium, as well as induce corrections to the
quantum-regression theorem. By considering the extra degrees of freedom as a
discrete manifold of energy levels, strong non-exponential behavior can
arise, as for example power law and stretched exponential decays.
\end{abstract}

\maketitle

\section{Introduction}

Irreversible, dissipative quantum dynamics (such as of an open system
embedded in an environment of uncontrolled degrees of freedoms) differs
drastically from reversible dynamics described by a unitary time-evolution
operator \cite{zurek,hanggi,weiss,leggett,alicki,nielsen,carmichael}. An
exact description of the dissipative dynamics can be given in the projector
formalism \cite{haake1}, which results in a master equation for the reduced
density matrix. In most cases, analytic progress can only be made under the
Markovian hypothesis, which requires that correlations between the system
and the environment decay faster than the characteristic inverse dissipation
rate $1/\gamma $. For weak coupling, Lindblad equations can be derived which
provide completely positive mapping of the density matrix from initial to
final conditions; the underlying quantum dynamical semi-group can also be
motivated from assumptions that are independent of the weak-coupling
requirement \cite{alicki,nielsen}. A cornerstone of this framework is the
quantum-regression theorem (QRT) \cite{lax,carmichael}, which relates
multiple-time correlation functions to single-time expectation values.
Feasible exact master equations beyond the Markovian approximation are known
when a spin \cite{hanggi,weiss,leggett} or a harmonic oscillator \cite{haake}
is embedded in a bosonic bath. Also, transient deviations from Markovian
behavior on times shorter than $1/\gamma $ are well understood \cite%
{suarez,gnutzmann,gaspard}. In general, however, only few general results
and manageable models are known for non-Markovian dynamics beyond the
transient regime \cite{barnett,wilkie,budini,cresser,lidar,chebotarev}.

The few recent results about strong non-Markovian effects in quantum master
equations were obtained in the context of complex environments \cite{wilkie}%
, continuous time quantum random walks \cite{budini}, stochastic
Hamiltonians \cite{cresser}, and continuous measurement theory \cite{lidar},
as well as in more mathematical settings \cite{chebotarev}. In general,
conditions for a QRT for non-Markovian dynamics have not been established in
the past.

In this paper, we present a new framework for the characterization of
non-Markovian quantum system dynamics. We show that long-time non-Markovian
effects naturally occur in composite environments where the
system-to-reservoir coupling strength depends on additional quantum degrees
of freedom. The non-Markovian behavior arises even when the reservoir itself
can be described in a Markovian approximation because the mediating degrees
of freedom become entangled with the system degrees of freedom. These
effects can persist for arbitrary long times, far beyond the transient
regime. When the mediating degrees of freedom are eliminated, the dynamics
can be rewritten as the statistical superposition of conventional master
equations with random dissipation rate. For weak coupling we obtain a class
of non-Markovian Lindblad equations which preserve the complete positivity
of the solution map known from the Markovian case. However, we find that the
QRT in general is not fulfilled, the only exception being the approach to a
stationary limit which is independent of the random dissipation rate. 
We also present an effective approximation to the system dynamics, which 
facilitates the comparison with previous results for non-Markovian dynamics.
As an illustrative example we analyze the
non-Markovian dispersive dynamics of a two level system. By assuming as the
extra system a manifold of quantum levels, we demonstrate the possibility of
getting strong non-exponential decays.

\section{Composite Environments}

We start from a full microscopic description, considering a system S that
interact with a composite bipartite reservoir consisting in a bath B
endowed with extra unobserved degrees of freedom U, which also participate
in the system-environment interaction. The total Hamiltonian reads%
\begin{equation}
H_{T}=H_{S}+(H_{U}+H_{B})+\lambda H_{I}  \label{hamiltonian}
\end{equation}%
with\ the tripartite interaction 
\begin{equation}
H_{I}=q_{S}\otimes (Q_{U}\otimes Q_{B}).  \label{eq:int}
\end{equation}%
The identity $\lambda H_{I}=(\lambda Q_{U})\otimes (q_{S}\otimes Q_{B})$
implies that the operator $Q_{U}$ sets the system-reservoir coupling
strength. For simplicity we assume that $Q_{U}$ is a constant of motion%
\begin{equation}
\lbrack H_{U},Q_{U}]=0.
\end{equation}%
Clearly, this assumption remains valid when the dynamics of $Q_{U}$ is slower than the dissipative relaxation \cite{footnote0}.

\subsection*{Reformulation in terms of a random interaction parameter
strength}

The dynamical evolution of the total density matrix $\rho _{T}(t)$ is given
by 
\begin{equation}
\rho _{T}(t)=\exp [\mathcal{L}_{T}\ t]\rho _{T}(0),
\end{equation}%
where $\mathcal{L}_{T}[\bullet ]=(-i/\hbar )[H_{T},\bullet ]$ is the total
Liouville superoperator. In order to relate these dynamics to conventional
dynamics in environments with fixed coupling strength, we eliminate the
unobserved degrees of freedom $Q_{U}$ for usual factorizing initial
conditions $\rho _{T}(0)=\rho _{SB}(0)\otimes \rho _{U}(0)$ \cite{footnote1}%
. The reduced density matrix $\rho _{SB}(t)=\mathrm{Tr}_{U}[\rho _{T}(t)]$
of the system S and the bath B is then given by 
\begin{equation}
\rho _{SB}(t)=\sum_{R}P_{R}\ \exp [(\mathcal{L}_{H}+\mathcal{L}_{B}+\lambda
_{R}\mathcal{L}_{I})t]\rho _{SB}(0),  \label{rho_SB}
\end{equation}%
where $\mathcal{L}_{H}$ and $\mathcal{L}_{B}$ are the Liouville operators of
the system and the bath respectively, and $\mathcal{L}_{I}[\bullet
]=(-i/\hbar )[(q_{S}\otimes Q_{B}),\bullet ]$. The index R runs over the
eigenstates $\left\vert R\right\rangle $ of $H_{U}$. We introduced the
probabilities 
\begin{equation}
P_{R}=\left\langle R\right\vert \rho _{U}(0)\left\vert R\right\rangle
\end{equation}%
and the weighted coupling strengths 
\begin{equation}
\lambda _{R}=\lambda \left\langle R\right\vert Q_{U}\left\vert
R\right\rangle .
\end{equation}%
Indeed, Eq.~(\ref{rho_SB}) can be interpreted as a statistical average $\rho
_{SB}(t)=\left\langle \rho _{SB}^{R}(t)\right\rangle $ over solutions $\rho
_{SB}^{R}(t)$ corresponding to a Hamiltonian 
\begin{equation}
H_{T}^{\prime }=H_{S}+H_{B}+\lambda _{R}\,q_{S}\otimes Q_{B}
\end{equation}%
with fixed interaction parameter $\lambda _{R}$. Each solution $\rho
_{SB}^{R}(t)$ participates with probability $P_{R}$. It follows that the
reduced system density matrix $\rho _{S}(t)=\mathrm{Tr}_{B}[\rho _{SB}(t)]$
can be obtained from the reduced density matrices $\rho _{R}(t)=\mathrm{Tr}%
_{B}[\rho _{SB}^{R}(t)]$ by a similar average 
\begin{equation}
\rho _{S}(t)=\sum_{R}P_{R}\,\rho _{R}(t)\equiv \langle \rho _{R}(t)\rangle .
\label{rho_SB2}
\end{equation}

The random coupling formulation (\ref{rho_SB2}) of the system dynamics
allows to incorporate the previous knowledge about dissipative systems with
fixed coupling strength \cite%
{zurek,hanggi,weiss,leggett,alicki,nielsen,haake1,carmichael,lax,haake,barnett,suarez,gnutzmann,gaspard}
(it also may form the basis for efficient numerical simulations). For
instance, if the map $\rho _{R}(0)\rightarrow \rho _{R}(t)$ is completely
positive then this is inherited by the map $\rho _{S}(0)\rightarrow \rho
_{S}(t)$.

In the rest of the paper, we will use the random description to make further
analytical progress for the case that the evolution of $\rho _{R}(t)$ can be
described by a Markovian Lindblad equation \cite{alicki}.

\section{Non-Markovian Lindblad equations}

When the correlation times of the bath B are the shortest time scale, a
Markovian approximation applies. For factorizing initial conditions of the
total density matrix, $\rho _{SB}(0)=\rho _{S}(0)\otimes \rho _{B}(0)$, and
weak coupling of S and B ($\lambda _{R}\ll 1$), one then can describe the
evolution of the reduced density matrix $\rho _{R}(t)$ by a Lindblad
equation \cite{alicki} 
\begin{equation}
\frac{d\rho _{R}(t)}{dt}=\mathcal{L}_{H}[\rho _{R}(t)]+\gamma _{R}\mathcal{L}%
[\rho _{R}(t)].  \label{lindblad}
\end{equation}%
The random dissipation rate is given by 
\begin{equation}
\gamma _{R}=\gamma \left\langle R\right\vert Q_{U}^{2}\left\vert
R\right\rangle ,
\end{equation}%
where $\gamma $ is determined by the spectral density of the environment
evaluated at a characteristic frequency of the system. The Lindblad
superoperator reads 
\begin{equation}
\mathcal{L}[\bullet ]=\frac{1}{2}\sum_{\alpha }([V_{\alpha },\bullet
V_{\alpha }^{\dagger }]+[V_{\alpha }\bullet ,V_{\alpha }^{\dagger }]),
\end{equation}%
with the operators $\{V_{\alpha }\}$ acting on the Hilbert space of S.

From this description, it follows that the evolution of the reduced density
matrix $\rho _{S}(t)=\langle \rho _{R}(t)\rangle $ is non-Markovian. In
fact, note that due to the statistical correlation between $\gamma _{R}$ and 
$\rho _{R}(t),$ the average of Eq.~(\ref{lindblad}) can not be written as a
evolution which is local in time%
\begin{equation}
\frac{d\rho _{S}(t)}{dt}\neq \{\mathcal{L}_{H}+\mathcal{L}^{\prime }\}[\rho
_{S}(t)],
\end{equation}%
with $\mathcal{L}^{\prime }$ being some extra Lindblad superoperator. Then,
for obtaining the corresponding non-Markovian master equation, we first
write the average Eq.~(\ref{rho_SB2}) in the Laplace domain as%
\begin{equation}
\rho _{S}(u)=\left\langle \frac{1}{u-(\mathcal{L}_{H}+\mathcal{L}_{R})}%
\right\rangle \rho _{S}(0)\equiv \left\langle G_{R}(u)\right\rangle \rho
_{S}(0),  \label{rholaplace}
\end{equation}%
where $\mathcal{L}_{R}=\gamma _{R}\mathcal{L}$ and $u$ is the Laplace
variable. In order to cast this expression into a deterministic closed
evolution equation we have to interchange the average over the random
dissipation rate and the operator-inverse in the definition of $G_{R}$. We
employ the identity%
\begin{equation}
\rho _{S}(u)=\frac{1}{\langle G_{R}(u)[u-(\mathcal{L}_{H}+\mathcal{L}%
_{R})]\rangle }\left\langle G_{R}(u)\right\rangle \rho _{S}(0),
\end{equation}%
and define a deterministic superoperator $\mathbb{L}$ by demanding 
\begin{equation}
\rho _{S}(u)=\frac{1}{u-[\mathcal{L}_{H}+\mathbb{L}(u)]}\;\rho _{S}(0).
\end{equation}%
The superoperator $\mathbb{L}$ is then determined by the condition 
\begin{equation}
\langle G_{R}(u)\mathcal{L}_{R}\rangle =\langle G_{R}(u)\rangle \mathbb{L}(u)
\label{memory}
\end{equation}%
in the Laplace domain, and fulfills 
\begin{equation}
\langle G_{R}(t)\mathcal{L}_{R}\rangle =\int_{0}^{t}d\tau \,\langle
G_{R}(t-\tau )\rangle \mathbb{L}(\tau )
\end{equation}%
in the time domain \cite{footnote3}.

The consequence of this procedure is a deterministic, closed, non-Markovian
evolution equation of the reduced density matrix, 
\begin{equation}
\frac{d\rho _{S}(t)}{dt}=\mathcal{L}_{H}[\rho _{S}(t)]+\int_{0}^{t}d\tau \,%
\mathbb{L}(t-\tau )[\rho _{S}(\tau )].  \label{kernel}
\end{equation}%
This equation has arbitrary long memory compared to the mean dissipation
rate $\langle \gamma _{R}\rangle $. By construction from the average of Eq.~(%
\ref{lindblad}), the solution map $\rho _{S}(0)\rightarrow \rho _{S}(t)$ of
Eq.~(\ref{kernel}) is completely positive.

\subsection{Quantum-Regression Theorem}

For Markovian Lindblad equations the QRT \cite{lax,carmichael} provides
direct relations between expectation values of system observable and their
correlation functions. We now use Eq.~(\ref{kernel}) to show that the
theorem cannot be taken for granted for composite environments. Let us
introduce a complete set of operators $\{A_{\mu }\}$ of the system,
collected into a vector $\mathbf{A}$, and consider the expectation values 
\begin{equation}
\overline{\mathbf{A}(t)}\equiv \mathrm{Tr}_{SUB}[\mathbf{A}(t)\rho _{T}(0)],
\end{equation}%
as well as the correlation functions 
\begin{equation}
\overline{S(t)\mathbf{A}(t+\tau )}\equiv \mathrm{Tr}_{SUB}[S(t)\mathbf{A}%
(t+\tau )\rho _{T}(0)],
\end{equation}%
where $S$ is an arbitrary operator for the system. The time dependence of
the operators refers to a Heisenberg representation with respect to the
total Hamiltonian (\ref{hamiltonian}).

Based on the random formulation (\ref{rho_SB2},\ref{lindblad}) of the
dynamics, the expectation values and correlation functions can be written as
averages over the dissipation rate, 
\begin{subequations}
\begin{eqnarray}
\overline{\mathbf{A}(t)} &=&\left\langle \mathrm{Tr}_{S}[\mathbf{A}\rho
_{R}(t)]\right\rangle \equiv \langle \overline{\mathbf{A}(t)}_{R}\rangle , \\
\overline{S(t)\mathbf{A}(t+\tau )} &=&\langle \mathrm{Tr}_{S}\{\mathbf{A}e^{(%
\mathcal{L}_{H}+\mathcal{L}_{R})\tau }[\rho _{R}(t)S]\}\rangle  \notag \\
&\equiv &\langle \overline{S(t)\mathbf{A}(t+\tau )}_{R}\rangle .
\label{correlation}
\end{eqnarray}%
The expressions deliver evolution equations 
\end{subequations}
\begin{subequations}
\begin{eqnarray}
\frac{d}{dt}\overline{\mathbf{A}(t)} &=&\langle \mathbf{\hat{M}}_{R}%
\overline{\mathbf{A}(t)}_{R}\rangle ,  \label{onepoint} \\
\frac{d}{d\tau }\overline{S(t)\mathbf{A}(t+\tau )} &=&\langle \mathbf{\hat{M}%
}_{R}\overline{S(t)\mathbf{A}(t+\tau )}_{R}\rangle ,  \label{twopoint}
\end{eqnarray}%
where the matrix $\mathbf{\hat{M}}_{R}$ acts on the indices of $\mathbf{A}$
and is defined by the condition 
\end{subequations}
\begin{equation}
\mathrm{Tr}_{S}\{\mathbf{A}(\mathcal{L}_{H}+\mathcal{L}_{R})[S]\}=\mathbf{%
\hat{M}}_{R}\mathrm{Tr}_{S}\{\mathbf{A}S\}.  \label{eq:m}
\end{equation}

When $\gamma_R$ is fixed, the evolution equations (\ref{onepoint}) for
expectation values and (\ref{twopoint}) for correlation functions are
identical, which recovers the QRT for Markovian dynamics. In the
non-Markovian case, however, both equations still involve the average over
the dissipation rate.

By using the same procedure as for the density matrix, we can transform Eq.~(%
\ref{onepoint}) into a closed deterministic evolution equation, 
\begin{equation}
\frac{d}{dt}\overline{\mathbf{A}(t)}=-\int_{0}^{t}d\tau \mathbb{\hat{M}}%
(t-t^{\prime })\overline{\mathbf{A}(t^{\prime })}.  \label{medio}
\end{equation}%
The deterministic kernel matrix $\mathbb{\hat{M}}(t)$ fulfills an equation
similar to Eq.~(\ref{memory}), but written in terms of $\mathbf{\hat{M}}_{R}$
and its corresponding propagator.

Equation (\ref{twopoint}) has the same structure as Eq.~(\ref{onepoint}),
but in the remaining average over the dissipation rate we are confronted
with a subtlety. While Eq.~(\ref{onepoint}) is defined with initial
conditions fixed at $t=0$, Eq.~(\ref{twopoint}) gives the solution with
initial condition $\overline{S(t)\mathbf{A}(t)}_{R}$ at finite time $t$.
From the definitions in Eqs.\ (\ref{correlation}) and (\ref{eq:m}) we find
statistical correlations between $\mathbf{\hat{M}}_{R}$ and $\overline{S(t)%
\mathbf{A}(t)}_{R}$, which both depend on $\gamma _{R}$. Dynamically these
correlations can be understood by realizing that $\overline{S(t)\mathbf{A}(t)%
}_{R}$ is of the form of a single-time expectation value and hence evolves
according to Eq.~(\ref{onepoint}). We still can disentangle the average over
the dissipation rate by the procedure employed for the density matrix and
the expectation value, but instead of a homogeneous equation of the form (%
\ref{medio}) obtain an inhomogeneous equation 
\begin{eqnarray}
\frac{d}{d\tau }\overline{S(t)\mathbf{A}(t+\tau )} &=&-\int_{0}^{\tau
}dt^{\prime }\ \mathbb{\hat{M}}(\tau -t^{\prime })\overline{S(t)\mathbf{A}%
(t+t^{\prime })}  \notag \\
&&+\mathbf{I}(t,\tau ),  \label{inhomogeneo}
\end{eqnarray}%
where the term $\mathbf{I}(t,\tau )$ accounts for the correlations.

The QRT is fulfilled when the inhomogeneity $\mathbf{I}(t,\tau )$ vanishes,
as is the case for Markovian dynamics where the average over the dissipation
rate is absent. Equation (\ref{correlation}) implies that the inhomogeneity
dies out in the long-time limit if the asymptotic state $\rho _{R}(\infty )$
does not depend on $\gamma _{R}$; the QRT is then asymptotically valid.
However, if the asymptotic state $\rho _{R}(\infty )$ depends on $\gamma
_{R} $ the inhomogeneous term will contribute at all times, even in the
asymptotic regime, and the QRT is invalidated.

\subsection{Effective Approximation}

In order to obtain a general characterization of the dynamics, we introduce
the following approximation. In Eq.~(\ref{memory}) we discard the dependence
introduced by the Lindblad superoperator $\mathcal{L}$ in the propagator $%
G_{R}(u)$, i.e., $\mathcal{L}_{R}\rightarrow -\gamma _{R}$I. Thus, we can
write the approximated equation as
\begin{equation}
\left\langle \frac{\gamma _{R}}{u-\mathcal{L}_{H}+\gamma _{R}}\right\rangle 
\mathcal{L}\approx \left\langle \frac{1}{u-\mathcal{L}_{H}+\gamma _{R}}%
\right\rangle \mathbb{L}(u),
\end{equation}%
which is solved by%
\begin{equation}
\mathbb{L}(u)\simeq K(u-\mathcal{L}_{H})\mathcal{L},
\end{equation}%
with the function%
\begin{equation}
K(u)=\left\langle \frac{\gamma _{R}}{u+\gamma _{R}}\right\rangle
\left\langle \frac{1}{u+\gamma _{R}}\right\rangle ^{-1}.  \label{K(u)}
\end{equation}%
From here, the density matrix evolution reads%
\begin{equation}
\frac{d\rho _{S}(t)}{dt}\simeq \mathcal{L}_{H}[\rho
_{S}(t)]+\int_{0}^{t}d\tau K(t-\tau )e^{(t-\tau )\mathcal{L}_{H}}\mathcal{L}%
[\rho _{S}(\tau )].  \label{aproximada}
\end{equation}%
In this approximation all information about the extra system $U$ is encoded
in the kernel $K(u)$, which is defined by Eq.~(\ref{K(u)}). While this
approximation is not controlled, it is clearly useful for characterization
of the possible non-Markovian effects. The solution of Eq.~(\ref{aproximada}%
) will differ from the exact solution of Eq.~(\ref{kernel}) only through
small time dependent corrections, of order 1, of the decay rate
parameters. On the other hand, we notice that the structure of the evolution
Eq.~(\ref{aproximada}) is similar to that found in Ref. \cite{lidar} in the
context of a continuous measurement theory.

\subsection{Stochastic state representation}

In the previous approximation, it is not clear whether the final evolution
guarantees the completely positive condition. Here, by introducing a
stochastic representation of the dynamics, we proof that indeed this
condition is preserved.

In the Laplace domain Eq.~(\ref{aproximada}) reads%
\begin{equation}
u\rho _{S}(u)-\rho _{S}(0)=\{\mathcal{L}_{H}+K(u-\mathcal{L}_{H})\mathcal{L}%
\}\rho _{S}(u).
\end{equation}%
By assuming that $\mathcal{L}=\mathcal{E}-$I \cite{footnote4}, with $%
\mathcal{E}[\bullet ]=\sum_{\alpha }V_{\alpha }\bullet V_{\alpha }^{\dagger
},$ we arrive at 
\begin{equation}
\rho _{S}(u)=\left\{ \frac{1}{\text{I}-w(u-\mathcal{L}_{H})\mathcal{E}}%
\right\} \frac{1-w(u-\mathcal{L}_{H})}{u-\mathcal{L}_{H}}\rho _{S}(0),
\end{equation}%
where we have introduced the function%
\begin{equation}
w(u)=\left\langle \frac{\gamma _{R}}{u+\gamma _{R}}\right\rangle .
\label{wait}
\end{equation}%
From here, it is possible to obtain the formal solution%
\begin{eqnarray}
\rho _{S}(t) &=&P_{0}(t)e^{t\mathcal{L}_{H}}\rho _{S}(0)  \label{average} \\
&&+\int_{0}^{t}d\tau \,w(t-\tau )e^{(t-\tau )\mathcal{L}_{H}}\mathcal{E}%
[\rho _{S}(\tau )].  \notag
\end{eqnarray}%
where $P_{0}(u)=[1-w(u)]/u=\left\langle (u+\gamma _{R})^{-1}\right\rangle .$
This equation have a clear stochastic interpretation. It corresponds to an
average of a stochastic density matrix $\rho _{st}(t)$ whose evolution
consists in the application, at random times, of the superoperator $\mathcal{%
E}$, while in the intermediate intervals the system state evolves with its
unitary evolution $\exp [t\mathcal{L}_{H}]$. The statistics of the random
times is dictated by $w(\tau )$, which can be interpreted as a waiting time
distribution, i.e., it is the probability density for an interval $\tau $
between consecutive applications of $\mathcal{E}$. In correspondence, $%
P_{0}(t)=1-\int_{0}^{t}d\tau w(\tau )$ is the survival probability
associated with $w(\tau )$. Then, the first term in Eq.~(\ref{average})
represents realizations without any application of the superoperator $%
\mathcal{E}$, while the integral term accounts for all other realizations.
Thus, we can write $\rho _{S}(t)=\left\langle \left\langle \rho
_{st}(t)\right\rangle \right\rangle $, where $\left\langle \left\langle
\cdots \right\rangle \right\rangle $ denotes the average over the random
times at which $\mathcal{E}$ is applied. As each realization preserves the
complete positivity, this property is also present in the averaged
evolution.

The previous stochastic framework allows us to clarify the role of the kernel $%
K(t)$. This follows after introducing the sprinkling distribution \cite%
{CoolingBook} $f(t)=w(t)\theta (t)+\int_{0}^{t}w(t-\tau )f(\tau ),$ where $%
\theta (t)$ is the step function \cite{footnote5}. From its definition, $%
f(t) $ gives the probability density for an event at time $t$, disregarding
the possibility of extra events in $(0,t)$. In the Laplace domain it reads $%
f(u)=w(u)/[1-w(u)]$. From here it is simple to get the relation $%
K(t)=df(t)/dt$, which defines the kernel as the rate of the sprinkling
distribution.

When the unitary dynamics commutate with the action of the superoperator $%
\mathcal{E}$, in an interaction representation the stochastic dynamics
reduces to that presented in Ref.~\cite{budini}. On the other hand, a
similar stochastic interpretation may be proposed for the exact evolution
Eq.~(\ref{kernel}), involving many renewal processes. Nevertheless, their
specific form depends on the details of each problem.

It is interesting to note that expressions similar to Eq.~(\ref{average})
arise in the context of the micromaser theory \cite{zoller,pickles,herzog}. This
system consists in an electromagnetic cavity that is continuously pumped
with excited atoms. In our scheme, the waiting time distribution $w(t)$ can
be associated with a non-Poissonian pump statistic, while the superoperator $%
\mathcal{E}$ with the transformation produced in the cavity field by the
passage of each atom. Therefore, the extra system $U$ can be associated with
the pump degrees of freedom. This comparison enlightens the dynamical origin
of the non-Markovian effects.

\subsection{Dephasing of a two-level system}

As an illustrative example of our results we consider a two-level system,
described by the Hamiltonian $H_{S}=(1/2)\hbar \omega _{A}\sigma _{z}$,
where $\sigma _{z}$ is the Pauli $z$-matrix, which is weakly connected to a
composite environment. The reservoir B is described by the dispersive
Lindblad operator%
\begin{equation}
\mathcal{L}[\bullet ]=\frac{1}{2}([\sigma _{z}\bullet ,\sigma _{z}]+[\sigma
_{z},\bullet \sigma _{z}]),
\end{equation}%
and the mediating part U is described by an arbitrary set ${\{\gamma
_{R},P_{R}\}}$ of dissipation rates and weights.

The evolution of the density matrix is given by the non-Markovian Lindblad
equation (\ref{kernel}). From Eq.~(\ref{memory}) we find the superoperator $%
\mathbb{L}(u)=K(u-\mathcal{L}_{H})\mathcal{L}$, with the kernel $K(u)$
defined in Eq.~(\ref{K(u)}). Thus, in this case, the evolution Eq.~(\ref%
{aproximada}) is exact.

The density matrix solution can be easily found in an interaction
representation with respect to the system Hamiltonian. We find the
completely positive map 
\begin{subequations}
\begin{eqnarray}
\rho _{S}(t) &=&g_{+}(t)\rho _{S}(0)+g_{-}(t)\sigma _{z}\rho _{S}(0)\sigma
_{z},  \label{sol1} \\
g_{\pm }(t) &=&\frac{1}{2}[1\pm P_{0}(t)].  \label{sol2}
\end{eqnarray}%
The function $P_{0}(t)=\sum_{R}P_{R}\exp [-\gamma _{R}t]$ is the survival
probability defined previously. Depending on the distribution of the
dissipation rate, arbitrary forms of the decay can be obtained from this
average over exponential functions. Hence the non-Markovian behavior can be
observed in the relaxation of the density matrix to the stationary state.

Let us now illustrate the consequences for the QRT, hence, the expectation
values and correlators of the vector of Pauli operators $\mathbf{A}\equiv
\{\sigma _{x},\sigma _{y},\sigma _{z},I\}$. For the expectation value (in
the interaction representation) we find 
\end{subequations}
\begin{equation}
\overline{\mathbf{A}(t)}=\mathbb{\hat{G}}(t)\overline{\mathbf{A}(0)},
\end{equation}%
where the matrix propagator reads $\mathbb{\hat{G}}(t)=\mathrm{diag}%
\{P_{0}(t),P_{0}(t),1,1\}$. Consistently with Eq.~(\ref{inhomogeneo}), the
correlation functions 
\begin{equation}
\overline{S(t)\mathbf{A}(t+\tau )}=\mathbb{\hat{G}}(\tau )\overline{S(t)%
\mathbf{A}(t)}+\mathbf{I}_{0}h(t,\tau ).
\end{equation}%
feature an extra inhomogeneity, which is given by $\mathbf{I}_{0}=\mathbb{\ 
\hat{D}}\mathrm{Tr}_{S}\{[\rho _{S}(0)-\rho _{S}(\infty )]S\mathbf{A}\}$,
with the matrix $\mathbb{\hat{D}}=\mathrm{diag}\{1,1,0,0\}$, and the
function $h(t,\tau )=P_{0}(t+\tau )-P_{0}(t)P_{0}(\tau )$. In the Markovian
case $h$ vanishes since the survival probability $P_{0}(t)$ is an
exponential, and the QRT is valid for all times. For non-Markovian dynamics,
the theorem is valid in the long-time asymptotic, since the inhomogeneous
term dies out as the equilibrium state $\rho _{S}(\infty )$ is attained
[consistently, in this example $\rho _{R}(\infty )$ is independent of the
dissipation rate $\gamma _{R}$].

\section{Entanglement with a discrete manifold of energy levels}

Up to now we have left unspecified the unobserved degree of freedom U, which
determine the set $\{\gamma _{R},P_{R}\}$. In this section we will analyze
the case in which U is defined by a discrete set of energy levels.

In the effective approximation, the properties of the unobserved degree of
freedom are introduced through the kernel $K(t)$, which is associated with
the renewal process defined by $w(t)=\sum_{R}P_{R}\;\gamma _{R}e^{-\gamma
_{R}t}$. This process has two characteristic times scales%
\begin{equation}
\langle \gamma \rangle =\sum_{R}P_{R}\;\gamma _{R},\ \ \ \ \ \ \ \ \langle
\tau \rangle =\sum_{R}P_{R}\;\gamma _{R}^{-1}.
\end{equation}%
When these constant are well defined, the average rate $\langle \gamma
\rangle $ defines an exponential decay of the waiting time distribution at
short times [$\langle \gamma \rangle t$ $<$ $1$], $\lim {}_{u\rightarrow
\infty }w(u)\approx \langle \gamma \rangle /(u+\langle \gamma \rangle ),$
and the average waiting time $\langle \tau \rangle =\int_{0}^{\infty
}dttw(t) $ characterizes an exponential decay of $w(t)$ in a long time
regime [$t$ $>$ $\langle \tau \rangle $], $\lim {}_{u\rightarrow
0}w(u)\approx 1-u\langle \tau \rangle .$ On the other hand, in terms of the
sprinkling distribution we get $\lim {}_{t\rightarrow 0^{+}}f(t)=\langle
\gamma \rangle $ and $\lim {}_{t\rightarrow \infty }f(t)=1/\langle \tau
\rangle .$

\subsection{Entanglement with a two-state system}

First we assume that the unobserved system U is a two level system. The
waiting time distribution then reads $w(t)=P_{\uparrow }\ \gamma _{\uparrow
}e^{-\gamma _{\uparrow }t}+P_{\downarrow }\ \gamma _{\downarrow }e^{-\gamma
_{\downarrow }t},$ with the condition $P_{\uparrow }+P_{\downarrow }=1$ and
arbitrary rates $\gamma _{\uparrow /\downarrow }$. Introducing the rates $%
\eta =P_{\uparrow }\gamma _{\downarrow }+P_{\downarrow }\gamma _{\uparrow }$
and $\beta =[\langle \gamma ^{2}\rangle -\langle \gamma \rangle
^{2}]/\langle \gamma \rangle $, the Laplace transform $w(u)$ can be written
as%
\begin{equation}
w(u)=\frac{\left\langle \gamma \right\rangle }{u+\left\langle \gamma
\right\rangle +\beta \sigma (u)},  \label{waittwo}
\end{equation}%
with $\sigma (u)=u/[u+\eta /(\langle \gamma \rangle \langle \tau \rangle )]$%
. The corresponding kernel Eq.~(\ref{K(u)}) results as%
\begin{equation}
K(u)=\frac{\langle \gamma \rangle }{1+\beta \sigma (u)/u},  \label{kerneltwo}
\end{equation}%
which in the time domain reads $K(t)=\langle \gamma \rangle \lbrack \delta
(t)-\beta e^{-\eta t}]$. We note that the fluctuation rate $\beta $ controls
departure from a Markov kernel. The sprinkling distribution results as $%
f(t)=\langle \gamma \rangle \theta (t)-[\langle \gamma \rangle -\langle \tau
\rangle ^{-1}](1-e^{-\eta t})$, where we have used $\eta \lbrack 1-(\langle
\gamma \rangle \langle \tau \rangle )^{-1}]=\beta $. As expected, $\langle
\gamma \rangle $ and $\langle \tau \rangle ^{-1}$ give respectively the
asymptotic values of $f(t)$ in the short and long time regimes, while $\eta $
gives the rate for the transition between these two regimes.

\subsection{Entanglement with a N-manifold of states}

Now we characterize the case in which the system U consists in a manifold of 
$N$ states [$0\leq R\leq N-1$] whose consecutive energy difference is
constant. We assume that the coupling strength of each level with the
system-bath-set decreases in an exponential way with the level energies, as
well as their stationary populations%
\begin{equation}
\gamma _{R}=\gamma \exp [-bR],\;\;\;\;\;\;P_{R}=\frac{(1-e^{-a})}{(1-e^{-aN})%
}\exp [-aR].
\end{equation}%
The constant $\gamma $ characterizes the Markovian decay of the system, and $%
b$ is a free dimensionless parameter that measure the strength of the
coupling between each states and the system-bath set, i.e., $%
Q_{U}=\sum_{R=0}^{N-1}\exp [-bR/2]|R\rangle \langle R|$. The second free
parameter $a$ measure the exponential decay of the populations. By taking $%
a=\hbar \omega _{0}/kT$, where $\hbar \omega _{0}$ is the difference of
energy between consecutive levels, the populations correspond to a thermal
distribution at temperature $T$.

The average rate reads%
\begin{equation}
\langle \gamma \rangle =\gamma \left( \frac{1-e^{-a}}{1-e^{-(a+b)}}\right)
\left( \frac{1-e^{-(a+b)N}}{1-e^{-aN}}\right) ,
\end{equation}%
from where the average waiting time follows immediately%
\begin{equation}
\langle \tau \rangle =\gamma ^{-1}\left( \frac{1-e^{-a}}{1-e^{-(a-b)}}%
\right) \left( \frac{1-e^{-(a-b)N}}{1-e^{-aN}}\right) .  \label{<tau>}
\end{equation}

In an intermediate regime, $\langle \gamma \rangle ^{-1}<t<\langle \tau
\rangle ^{-1}$, the corresponding waiting time distribution may present a 
\textit{power law behavior}. In fact, in the limit of $N\rightarrow \infty $%
, where the $N$-manifold states is equivalent to a thermal harmonic
oscillator in equilibrium at temperature $T$, it is possible to proof that,
after a time transient of order $1/\gamma $, the waiting time distribution
behaves as \cite{alemany}%
\begin{equation}
w(t)\approx 1/(\gamma t)^{1+\alpha },\;\;\;\;\;\;\;\;\;\alpha =a/b.
\end{equation}
For finite $N$, this behavior is also present, nevertheless the asymptotic
behavior changes to an exponential decay with rate $\langle \tau \rangle
^{-1} $.

For $a<b$ [$0<\alpha <1$] the behavior of $w(t)$ can be captured with a
simple analytical expression. Taking in account the results for the case $%
N=2 $, Eqs. (\ref{waittwo})-(\ref{kerneltwo}), we propose the complete
monotone function \cite{budini}%
\begin{equation}
w(u)=\frac{\left\langle \gamma \right\rangle }{u+\left\langle \gamma
\right\rangle +\beta ^{1-\alpha }\sigma _{\alpha }(u)},
\end{equation}%
where as before $\beta \approx \lbrack \langle \gamma ^{2}\rangle -\langle
\gamma \rangle ^{2}]/\langle \gamma \rangle $, and now $\sigma _{\alpha
}(u)=[(u+\gamma _{c})^{\alpha }-\gamma _{c}^{\alpha }]$. Using the relation $%
K(u)=w(u)/P_{0}(u)$, the kernel reads%
\begin{equation}
K(u)=\frac{\left\langle \gamma \right\rangle }{1+\beta ^{1-\alpha }\sigma
_{\alpha }(u)/u}.
\end{equation}%
From $\lim {}_{u\rightarrow 0}w(u)\approx 1-[u+\beta ^{1-\alpha }\sigma
_{\alpha }(u)]/\langle \gamma \rangle $, the asymptotic exponential decay of 
$w(t)$ with rate $\langle \tau \rangle ^{-1}$ can be fitted with the cutoff $%
\gamma _{c}$ after imposing the relation $\alpha (\beta /\gamma
_{c})^{1-\alpha }=\langle \gamma \rangle \langle \tau \rangle -1$. On the
other hand the constant $\beta $ fits the power law regime. In fact, in the
limit $N\rightarrow \infty $, the average waiting time Eq.~(\ref{<tau>}) is
infinite, which implies $\gamma _{c}=0$. Thus, $\lim {}_{u\rightarrow
0}w(u)\approx 1-\beta ^{1-\alpha }u^{\alpha }/\langle \gamma \rangle $,
implying a pure power law asymptotic behavior \cite{metzler}.

The previous analysis demonstrates that the transient behavior between the
short and asymptotic exponential decays of $w(t)$ is described by a power
law. As we have seen in the previous section, this behavior is in general
reflected by the system dynamics. On the other hand, we notice that when $%
N\rightarrow \infty $, by maintaining the average rate $\left\langle \gamma
\right\rangle $ fixed, the fluctuation rate $\beta $ reaches it maximum
value in the limit of both, strong coupling $b\ll 1$ and small populations
decay $a\ll 1$ (which can be also read as a high temperature limit). In this
case, after the transient $t<\left\langle \gamma \right\rangle ^{-1}$, the
waiting time distribution and the kernel can be approximated by the
expressions%
\begin{equation}
w(u)\approx \frac{A_{\alpha }}{A_{\alpha }+u^{\alpha }},\ \ \ \ \ \ \ \ \
K(u)\approx A_{\alpha }u^{1-\alpha },
\end{equation}%
with $A_{\alpha }\approx \left\langle \gamma \right\rangle /\beta ^{1-\alpha
}\approx \left\langle \gamma \right\rangle ^{\alpha }$. These expressions
correspond to a \textit{fractional derivative evolution} \cite{budini},
where stretched exponential and power law behavior arises jointly.

\section{Conclusions}

We have shown that a system in a composite bipartite environment (where the
system-to-reservoir coupling depends on other degrees of freedom) follows
non-Markovian dynamics even when the reservoir itself can be eliminated by a
Markovian approximation. The non-Markovian effects originate in the
entanglement of the system with the mediating degrees of freedom, and may
persist for arbitrary long times.

Our results are derived from a random rate reformulation of the dynamics in
the composite environment which allows to make full contact to the
established theory of dissipative systems with constant coupling. On this
basis, we formulated non-Markovian Lindblad equations which provide complete
positive mappings of the density matrix from initial to final conditions,
and identified conditions for the quantum-regression theorem. It should be
noted that the random rate formulation is not restricted to the Lindblad
master equations but can be applied to other master equations, including
exact master equations which may already include non-Markovian effects at
fixed coupling.

In an effective approximation all information about the extra degrees of
freedom is introduced by a memory kernel. The corresponding density matrix
evolution can be interpreted in terms of stochastic process in the system
Hilbert space. System decay behaviors ranging from stretched exponential to
power law can be obtained by taking the system U as a discrete manifold of
states.

The present formalism substantiates previous results on non-Markovian master
equations \cite{barnett,wilkie,budini,cresser,lidar,chebotarev}, and puts
them into the alternative and greater perspective of systems embedded in a
composite environment. Our motivation to study this kind of environment
arises from recent experiments on fluorescent single quantum dots \cite%
{nirmal,michler,brokmann,grigolini}, where non-Markovian effects on time
scales much larger than $1/\gamma $ were found. While the underlying
physical mechanisms are not completely clear, it has been argued \cite%
{kuno,schlegel} that the experimental results can only be recovered when one
accounts for additional degrees of freedom which modulate the dissipative
coupling. Thus, besides its theoretical interest in the context of strong
non-Markovian effects in open quantum systems, the characterization of the
dynamics induced by composite environments may be also of interest in those
experiments. On the other hand, we believe that the present results may be
useful in modeling the dynamics of open quantum systems embedded in complex
structured host environments. In fact, the tri-partite interaction
investigated by us naturally arises when one considers the dynamical effects
of a disordered condensed-matter environment on a system coupled to a (say,
Markovian) environment. The coupling strength of the system to the Markovian
environment is proportional to the density of states of the Markovian
environment at a characteristic frequency of the system; this density of
states, in turn, depends on the condensed-matter environment (e.g., the
disorder configuration, or the charging of trap states). Taking the dynamics
of the condensed-matter environment into account, the coupling strength then
becomes as a dynamical variable, leading to the multiplicative tri-partite
coupling that was considered in this paper.

In the description of condensed-matter systems, multipartite interactions
also frequently appear as a consequence of a suitable transformation (such
as the Fr{\"{o}}hlich transformation) \cite{wagner}. Such transformations
can be used to eliminate (at least to some order of a small parameter) the
time-dependence of a certain subsystem, which then takes on the
characteristics of our (stationary) unobserved degrees of freedom.


\begin{thebibliography}{99}
\bibitem{footnote0} For very slow dynamics of $Q_{U}$ (more slowly than the
dissipative relaxation rate) the system will adiabatically follow the
stationary state of Eq.~(\ref{kernel}).
\bibitem{footnote1} For non-factorizing initial conditions between S and U,
the evolution of the system can be obtained by using projector operator
techniques. In this case, a mapping onto a model with random dissipation
rate is also possible, but initial correlations between the system and the
dissipation rate have to be considered. This implies an extra inhomogeneous
term in the averaged evolution Eq.~(\ref{kernel}), similar to the one
encountered in Eq.~(\ref{inhomogeneo}).
\bibitem{footnote3} The typical solution for the superoperator $\mathbb{L}%
(u) $ is a sum of Lindblad superoperators, each one with its corresponding
kernel. For the analysis of equations with a similar structure see Refs.\ 
\cite{wilkie,budini,cresser,lidar}.
\bibitem{footnote4} In a general case, where $\mathcal{L}\neq \mathcal{E}-$%
I, with $\mathcal{E}$ a completely positive superoperator, the stochastic
dynamics can only be established in a formal way after introducing a limit
procedure \cite{budini}. In this case, the superoperator reads $\mathcal{E}%
[\rho ]=\{$I$+[e^{\kappa \mathcal{L}}-$I$]\}\rho $, where $\kappa $ must be
intended as a control parameter. Eq.~(\ref{aproximada}) is recuperated in
the limit in which simultaneously $\kappa \rightarrow 0$ and the number of
events by unit of time go to infinite, the last limit being controlled by
the sprinkling distribution $f(t)\rightarrow f(t)/\kappa $\ \cite%
{CoolingBook} associated to the waiting time distribution $w(t)$. Then,
while the stochastic dynamics can be formally extended to the general case,
it is not possible to implement it as a numerical method to solve the
corresponding master equation.
\bibitem{footnote5} Here the step function $\theta (t)$ is defined as $%
\theta (t)=1$ for $t>0$, and $\theta (t)=0$ for $t\leq 0$. With this
definition it follows that $f(t)|_{t=0}=0$.
\end{thebibliography}
\end{document}